\documentclass[sigplan,10pt]{acmart}
\AtBeginDocument{
  \providecommand\BibTeX{{
    \normalfont B\kern-0.5em{\scshape i\kern-0.25em b}\kern-0.8em\TeX}}}

\copyrightyear{2023}
\acmYear{2023}
\setcopyright{acmcopyright}\acmConference[EuroSys '23]{Eighteenth European
Conference on Computer Systems}{May 8--12, 2023}{Rome, Italy}
\acmBooktitle{Eighteenth European Conference on Computer Systems (EuroSys
'23), May 8--12, 2023, Rome, Italy}
\acmPrice{15.00}
\acmDOI{10.1145/3552326.3567495}
\acmISBN{978-1-4503-9487-1/23/05}

\usepackage{xspace}
\usepackage{url}
\usepackage{graphicx}
\usepackage{makecell}
\usepackage{threeparttable}
\usepackage{enumitem}
\usepackage{tikz}
\usepackage{hyperref}

\newcommand*\circled[1]{\tikz[baseline=(char.base)]{
            \node[shape=circle,draw,inner sep=0.5pt] (char) {#1};}}

\newcommand{\sysname}{\textsc{Rio}\xspace}
\newcommand{\fsname}{\textsc{RioFS}\xspace}
\newcommand{\horae}{\textsc{Horae}\xspace}
\newcommand{\horaefs}{\textsc{HoraeFS}\xspace}
\newcommand{\barrierfs}{BarrierFS\xspace}

\begin{document}

\title{\sysname: Order-Preserving and CPU-Efficient Remote Storage Access}
\date{}

\author{Xiaojian Liao, Zhe Yang, Jiwu Shu} 
\authornote{Jiwu Shu is the corresponding author (shujw@tsinghua.edu.cn).}
\affiliation{ 
 \institution{\textit{Department of Computer Science and Technology, Tsinghua University} \\ \textit{Beijing National Research Center for Information Science and Technology (BNRist)}}
 \country{}
 \city{}
}

\begin{abstract}
Modern NVMe SSDs and RDMA networks provide dramatically higher bandwidth and concurrency.
Existing networked storage systems (e.g., NVMe over Fabrics) fail to fully exploit these new devices due to inefficient storage ordering guarantees.
Severe synchronous execution for storage order in these systems stalls the CPU and I/O devices and lowers the CPU and I/O performance efficiency of the storage system.

We present \sysname, a new approach to the storage order of remote storage access.
The key insight in \sysname is that the layered design of the software stack, along with the concurrent and asynchronous network and storage devices, makes the storage stack conceptually similar to the CPU pipeline.
Inspired by the CPU pipeline that executes out-of-order and commits in-order, \sysname introduces the I/O pipeline that allows internal out-of-order and asynchronous execution for ordered write requests while offering intact external storage order to applications.
Together with merging consecutive ordered requests, these design decisions make for write throughput and CPU efficiency close to that of orderless requests.

We implement \sysname in Linux NVMe over RDMA stack, and further build a file system named \fsname atop \sysname.
Evaluations show that \sysname outperforms Linux NVMe over RDMA and a state-of-the-art storage stack named \horae by two orders of magnitude and 4.9$\times$ on average in terms of throughput of ordered write requests, respectively.
\fsname increases the throughput of RocksDB by 1.9$\times$ and 1.5$\times$ on average, against Ext4 and \horaefs, respectively.
\end{abstract}

\keywords{Storage Order, NVMe over Fabrics, Flash, File System, SSD}

\ccsdesc[500]{Information systems~Information storage systems}

\maketitle

\section{Introduction}
Remote storage access (i.e., accessing storage devices over the network) has become increasingly popular for modern cloud infrastructures and datacenters to share the enormous capacity and bandwidth of fast storage devices~\cite{flash-disaggregation-eurosys16,facebook-datacenter}.
Unlike legacy HDDs and SATA SSDs whose maximum bandwidth is less than 750~MB/s due to the interface limit, a commodity NVMe SSD provides nearly 7~GB/s bandwidth and 1.5 million IOPS~\cite{intel-p5800x}.
The speeds of an RDMA NIC have transitioned to 200~Gbps for fast data transfer among servers~\cite{Mellanox-connectx-6}.
These changes make CPU efficiency also a dominant factor in storage and network systems~\cite{max-atc21,asyncio-atc19,KVell-sosp19,ccNVMe-sosp21,i10-nsdi20,scalefs-sosp17,zjournal-atc21,scalexfs-fast22,mtcp-nsdi14,Linux-block-io}.
This paper targets storage order, which is the fundamental building block of storage consistency, and which often prevents reliable storage systems from exploiting these high-performance hardware devices.

Storage order indicates a certain persistence order of data blocks to storage media.
It is extensively used in storage consistency mechanisms (e.g., database transactions~\cite{MySQL,sqlite}, soft updates~\cite{soft-updates} and file system journaling~\cite{file-journaling,ext4}) to ensure deterministic and correct disk states despite a system crash.
For decades, traditional networked storage systems use a quite expensive approach to ensure storage order. 
The following ordered write requests can not be processed until preceding requests are completed and associate data blocks are durable (\S\ref{sec:background}).
This \textit{synchronous} approach, however, leaves NICs and SSDs underutilized and the CPU in an idle state.

We first seek state-of-the-art approaches from the local storage stack to see if similar designs can be applied to networked storage to mitigate the performance overhead.
Unfortunately, their approaches prevent the streamlined use of CPU, I/O devices, or both, thereby offering suboptimal CPU and I/O efficiency and making it difficult to scale to multiple servers.
For example, \horae~\cite{horae-osdi20}, a recently proposed order-preserving approach, introduces a dedicated control path for storage order.
However, the control path is synchronous and executed before the data path, which wastes considerable CPU cycles and further lowers the I/O throughput (\S\ref{sec:motivation}).

We observe that the layered design (e.g., the network and storage drivers) of the software stack, along with the concurrent and asynchronous network and storage devices, makes the storage stack conceptually similar to the CPU pipeline.
We thus introduce the I/O pipeline for ordered write requests (\S\ref{sec:design}).
The I/O pipeline adopts the out-of-order and asynchronous execution from the CPU pipeline.
It speculatively executes ordered write requests that target non-overlapping data blocks in parallel as if they were orderless requests.
As NICs or SSDs can be easily saturated by orderless requests, this \textit{asynchronous} approach fully exploits NICs and SSDs.
In addition, asynchronous execution lets CPUs continuously push I/O requests to the hardware, without being idle or switched out, thereby making more efficient use of CPUs.

In the I/O pipeline, since each layer of the storage stack can process multiple ordered write requests from different cores, asynchronous execution brings uncertainties that may violate the original ordering semantics or even sacrifice data persistence.
We introduce a series of techniques including \textit{in-order submission and completion} and leverage the \textit{in-order delivery} of the network protocol, to reduce temporary out-of-order execution, thus removing uncertainties and providing final intact storage order to applications.
Even if a crash occurs, our proposed \textit{asynchronous crash recovery} algorithm can quickly recover the system to an ordered state.
The key enabler of these techniques is a special structure called the \textit{ordering attribute}, which is an identity of each ordered write request and tracks neighboring ordered write requests.
It is embedded in the original request and carried throughout the storage stack.
Therefore, although being asynchronous, each ordered write request is able to collect the scattered ordering attributes and reconstruct the original storage order at any time using the aforementioned techniques.

We implement this design within \sysname, an order-preserving networked storage stack, with a set of modifications of Linux NVMe over RDMA stack (\S\ref{sec:imple}).
We further develop a file system called \fsname that uses the ordered block device of \sysname.
We evaluate \sysname and \fsname with two kinds of SSDs (i.e., flash and Optane SSDs), against Linux NVMe over RDMA stack and an NVMe-oF version of \horae~\cite{horae-osdi20} which is originally an order-preserving storage stack for local NVMe SSDs and is extended to support networked storage (\S\ref{sec:evaluation}).
We find that \sysname and \fsname perform significantly better than their counterparts.
The throughput and CPU efficiency of \sysname even come close to the orderless write requests.

In summary, we make the following contributions:
\begin{itemize}[noitemsep,topsep=0pt,parsep=0pt,partopsep=0pt]
        \item We conduct a study on existing methods for storage order and summarize three lessons for building a high-performance and order-preserving storage stack.
        \item We propose \sysname that achieves storage order, high performance and CPU efficiency  simultaneously.
        \item We implement and evaluate \sysname and \fsname, demonstrating significant performance and CPU efficiency improvement over state-of-the-art systems.
\end{itemize}

\section{Background and Related Work}\label{sec:background}

\subsection{Remote Storage Access}\label{sec:bg_storage_disacggregation}
The networked storage consists of two major components: the initiator and target.
The target represents both the software and hardware of a remote device.
It exposes local storage devices to remote initiators via the standard block storage protocol (e.g., NVMe~\cite{NVMe}) over a range of network fabrics (e.g., RDMA, TCP).
The initiator can thus access the remote block device as if using a local storage device.
Recently, NVMe-oF (NVMe over Fabrics) is introduced as an alternative to the traditional iSCSI~\cite{iSCSI} owing to its low protocol overhead and high parallelism.
A number of works~\cite{reflex-asplos17,i10-nsdi20,gimbal-sigcomm21,flash-disaggregation-eurosys16} have studied and improved the orderless I/O path of the networked storage.
To preserve storage order, they still rely on traditional synchronous execution as in the Linux NVMe-oF.
Our proposal for storage order is orthogonal to their designs.

\begin{figure}[t] 
	\centering 
	\includegraphics[width=\linewidth]{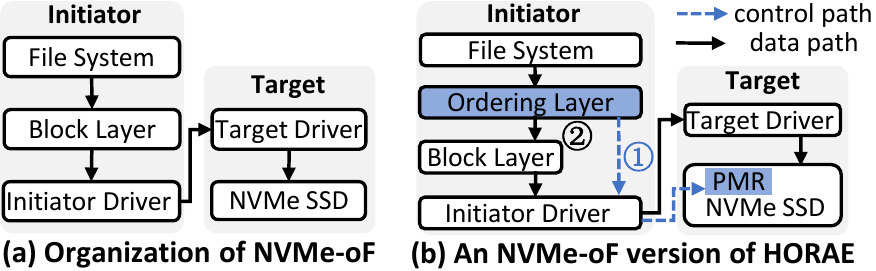} 
	\vspace{-10pt}
	\caption{\textbf{Background of NVMe-oF and \horae.}}
	\vspace{-10pt}
	\label{fig:background_nvme_rdma} 
\end{figure}

As our implementation is based on NVMe-oF, we present its details (Figure~\ref{fig:background_nvme_rdma}(a)).
In NVMe-oF, the file system, block layer and NVMe SSD are almost the same as the local NVMe over PCIe stack.
The major difference lies in the initiator and target drivers that control how I/O commands and data blocks are transferred over the network. 
Specifically, if the network fabric is RDMA, I/O commands that describe the source and destination addresses of data blocks and completion responses are transferred via two-sided RDMA SEND operations.
The data blocks are transferred by one-sided RDMA READ or WRITE operations. 
Note that one-sided operations bypass the target CPU, but two-sided operations require the target CPU to search and update RDMA queues.

\subsection{Storage Order}\label{sec:bg_stroage_order}
The storage order defines a certain order of data blocks to be persisted in the storage media.
Traditional Linux I/O stacks such as NVMe-oF do not offer storage ordering guarantees.
They use a synchronous execution approach in the file systems (e.g., synchronous transfer and \texttt{FLUSH} commands) or applications (e.g., \texttt{fsync}) to achieve storage order.
The cost of the traditional approach is expensive, and recent studies~\cite{optfs-sosp13,barrierfs-fast18,horae-osdi20,optr-atc19,scftl-osdi20,oafa-systor22} attempt to reduce the overhead of storage order for local SCSI and NVMe stacks.
In this section, we introduce Linux NVMe-oF and \horae~\cite{horae-osdi20} on NVMe stack, followed by discussing \barrierfs~\cite{barrierfs-fast18} and OptFS~\cite{optfs-sosp13} designed on the older SCSI stack.

Each layer of \textbf{NVMe-oF} is orderless.
For example, requests to different queues of the drivers can be executed out-of-order.
The NVMe SSD may freely re-order requests due to massive internal data parallelism.
Thus, to control storage order in NVMe-oF, the file system issues the next ordered write request only if associated data blocks of the preceding request are durable.
Specifically, for each ordered write request, the file system does not issue the next request until the request flows through the block layer, initiator and target drivers and reaches the NVMe SSD, and data blocks are ensured to be durable by a \texttt{FLUSH} command on the SSD.
The overhead of this approach which we call synchronous execution is severe~\cite{optfs-sosp13,nofs-fast12,barrierfs-fast18,horae-osdi20} in local storage and becomes worse in the networked storage (\S\ref{sec:motivation}).

\textbf{\horae} separates the storage ordering control from the request flow and provides a dedicated control path (Figure~\ref{fig:background_nvme_rdma}(b)).
The control path is used to ensure storage order first, and thus ordered write requests can be processed asynchronously and concurrently.
Specifically, in the control path, \horae stores ordering metadata that retains enough information to recover from an untimely crash to the persistent memory region (PMR)~\cite{nvme-pmr,NVMe-1.4} of the NVMe SSD.
The PMR is a region of general purpose read/write persistent memory of the SSD.
It is byte-addressable and can be accessed directly by CPU load and store instructions (MMIO operations), thus making the control path fast.
However, since control path operations are executed before the ordered write requests are serviced, the control path is synchronous and becomes a serialization bottleneck (\S\ref{sec:motivation}).

\textbf{\barrierfs} keeps each layer order-preserving.
For example, the block layer schedules ordered write requests in a FIFO fashion.
The SSD needs a barrier write command to understand the storage order and make data blocks durable in order.
This approach is overly strict to storage order and thus makes it difficult to extend the idea to support modern multi-queue hardware (e.g., NVMe SSD and RDMA NIC)~\cite{horae-osdi20}, multiple targets~\cite{oafa-systor22} and servers.
For example, to agree on a specific order, requests from different cores contend on the single hardware queue, which limits the multicore scalability.
SSDs are unable to communicate with each other, and thus keeping storage order among multiple targets is challenging.
As we will show in \sysname, intermediate storage order is not a necessity and can be relaxed.

\textbf{OptFS} introduces an optimistic approach that uses transaction checksums to detect ordering violations and perform further crash recovery.
This approach requires more CPU cycles to calculate and validate checksums.
Such an investment of CPU cycles is beneficial for HDDs as the speed gap between legacy storage devices and CPUs is so large.
However, for modern NVMe SSDs, CPU cycles are no longer a negligible part~\cite{asyncio-atc19,KVell-sosp19,max-atc21}.
A study~\cite{Vijay-thesis-optfs} from the same authors reveals that OptFS does not perform well on SSDs since the CRC32 checksum computation is a significant part of the total run-time. 
Furthermore, the transaction checksum is restricted to systems that place data blocks of ordered write requests in pre-determined locations (e.g., journaling).
Hence, it does not offer a lower-level ordered block device abstraction that can be randomly written and atop which many applications are built (e.g., BlueStore~\cite{bluestore}, KVell~\cite{KVell-sosp19}).
\section{Motivation}\label{sec:motivation}
In this section, we quantify and analyze the overhead of storage ordering guarantees of remote storage access.

\subsection{Motivation Experiments}
We perform experiments on both ordered and orderless write requests with Linux NVMe over RDMA and \horae~\cite{horae-osdi20}.
We extend \horae to NVMe over RDMA stack (details in \S\ref{sec:eval_setup}).
Since we do not have barrier-enabled storage and can not control the behavior of the NIC, we are unable to evaluate \barrierfs.
As OptFS is implemented in an old Linux with no support for NVMe and RDMA, we can not compare it by experiments. 
Other details of the testbed are described in \S\ref{sec:evaluation}.

\begin{figure}[t] 
	\centering 
	\includegraphics[width=\linewidth]{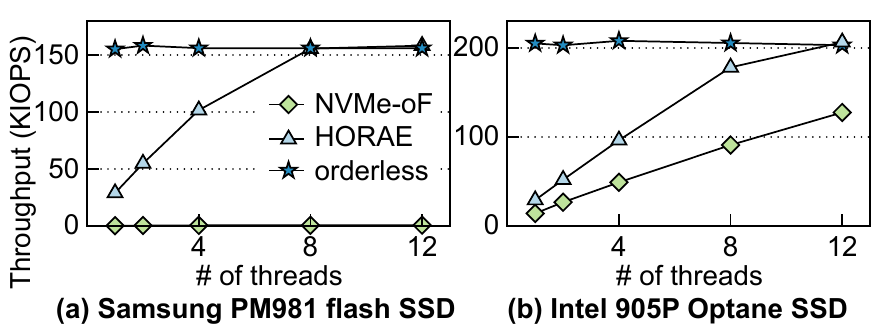} 
	\vspace{-10pt}
	\caption{\textbf{Motivation experiments.}
	\textit{NVMe-oF: NVMe over RDMA with ordering guarantees. orderless: NVMe over RDMA with no ordering guarantee.}} 
	\vspace{-10pt}
	\label{fig:motivation} 
\end{figure}

The test launches up to 12 threads, and each performs the following workload to a private SSD area independently.
Each thread issues an ordered write request that contains 2 continuous 4~KB data blocks, and then performs another 4~KB consecutive ordered write request.
We choose this workload as it simulates the write pattern of the metadata journaling widely used in storage systems.
Specifically, the first 2 data blocks represent the journal description and metadata blocks, and the final 4~KB block is the commit record.
Applications (e.g., MySQL) that require strong consistency and durability issue \texttt{fsync} to trigger the metadata journaling.
Figure~\ref{fig:motivation} plots the average throughput.
The gap between the orderless which does not guarantee storage order and other systems indicates the overhead of guaranteeing storage order.

We make two observations from Figure~\ref{fig:motivation}.
First, orderless write requests which are executed asynchronously saturate the bandwidth of both the flash and Optane SSD with only a single thread.
Second, Linux NVMe-oF (NVMe over RDMA) and \horae perform significantly worse than the orderless.
\horae needs more than 8 CPU cores to fully drive existing SSDs.
For storage arrays and newer and faster SSDs, it is expected that it needs more computation resources.
The results of Figure~\ref{fig:motivation} therefore indicate that the cost of storage ordering guarantees of existing approaches is expensive.

\subsection{Analysis and Lessons}\label{sec:analysis_lessons}
We examine the behaviors of the ordered I/O path and decompose the storage order overhead.
The ordered I/O path consists of three main parts: CPU execution on the software, data transfer over network and PCIe, and device execution in particular the hardware barrier instructions (e.g., \texttt{FLUSH}).
We analyze each part at length and summarize three lessons.

\noindent
\textbf{Lesson 1: alleviating the overhead of storage barrier instructions.}
The ordered NVMe-oF suffers from the classic barrier instruction (i.e., \texttt{FLUSH}).
On the flash SSD with a volatile write cache (Figure~\ref{fig:motivation}(a)), NVMe-oF issues a \texttt{FLUSH} command for each ordered request to ensure that preceding data blocks are durable.
The \texttt{FLUSH} is a synchronous activity and flushes nearly all content including data blocks and FTL mappings from the device's volatile cache to persistent flash memory.
\horae and the orderless remove the \texttt{FLUSH}.
Comparing NVMe-oF with \horae, we observe that the \texttt{FLUSH} is quite expensive and thus lowers the throughput dramatically.

\begin{figure}[t] 
	\centering 
	\includegraphics[width=\linewidth]{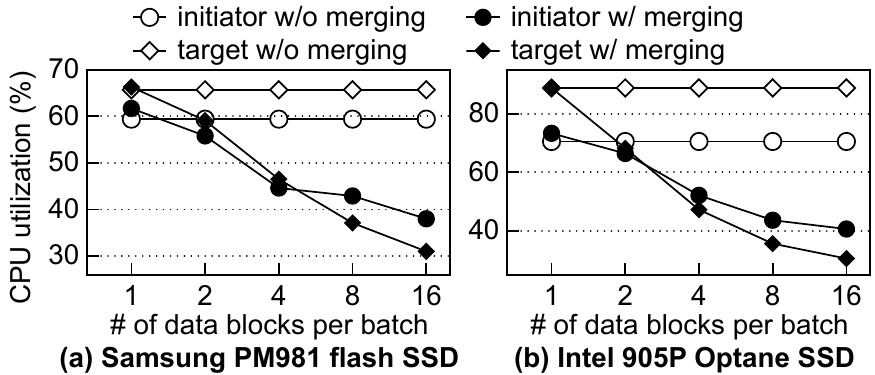} 
	\vspace{-10pt}
	\caption{\textbf{Motivation for merging consecutive data blocks.}
	\textit{Tested system: the orderless Linux NVMe over RDMA.}} 
	\vspace{-10pt}
	\label{fig:motivation_merging} 
\end{figure}

\noindent
\textbf{Lesson 2: making data transfer asynchronous.}
The Optane SSD in Figure~\ref{fig:motivation}(b) has a power loss protection technique (e.g., a non-volatile write cache).
Hence, the overhead of the \texttt{FLUSH} is marginal in this kind of SSDs.
Here, the dominant factor is data transfer over the network and PCIe bus.
The ordered NVMe-oF dispatches the next ordered write requests after the preceding request reaches the SSD.
This synchronous approach, however, leaves the NIC and SSD underutilized.

\horae separates the storage order from the request flow, and thus makes the transfer of data blocks asynchronous.
This approach allows more outstanding requests to be processed by NICs and SSDs.
Nonetheless, the control path is executed synchronously before the data path.
Comparing the orderless which transfers all data asynchronously with \horae (Figure~\ref{fig:motivation}), we find that the synchronous control path of \horae decreases the throughput significantly.

We dive into \horae's control path to understand the inefficiency.
The control path is essentially a faster and byte-addressable I/O path.
An ideal implementation of the control path in NVMe-oF is based on one-sided RDMA operations and PCIe peer-to-peer (P2P).
Specifically, the NIC can issue PCIe transactions on the PMR by P2P, bypassing the target CPU and memory.
The initiator driver first invokes an RDMA WRITE operation to dispatch the ordering metadata, and then issues an RDMA READ operation to ensure that the ordering metadata reaches PMR.
The latency of this ideal control path is expected to be larger than 4~$\mu$s, the raw hardware latency of a persistent RDMA write~\cite{rdma-nvm-atc21}.
Modern NVMe SSDs deliver a 4~KB I/O latency of sub-ten $\mu$s~\cite{asyncio-atc19,intel-905-optane} and the latency tends to decrease with newer PCIe 4.0 and 5.0 SSDs~\cite{intel-p5800x}.
As the latency of the SSD is comparable to the control path, the overhead of the synchronous control path is non-negligible.
In summary, to fully exploit the fast hardware devices, all data (including any control information) transfer over the network and PCIe should be asynchronous.

\noindent
\textbf{Lesson 3: reducing CPU cycles whenever possible.}
If the I/O stack alleviates the overhead of storage barrier instructions and makes data transfer asynchronous, the bottleneck will be shifted to the CPU overhead (i.e., CPU cycles spent on the I/O stack).
Specifically, in ordered NVMe-oF and \horae, a large fraction of CPU cycles are consumed at the device drivers, e.g., two-sided RDMA SEND operations.
Synchronous execution prevents the I/O stack from merging consecutive ordered write requests.
The I/O stack generates an I/O command for each request, and each I/O command requires many CPU cycles on RDMA and NVMe queues.
When ordered write requests become asynchronous as the orderless, they can be merged to reduce the CPU overhead.
We elaborate on this by Figure~\ref{fig:motivation_merging}.

Figure~\ref{fig:motivation_merging} presents the CPU overhead collected by the \texttt{top} command when we test the orderless NVMe-oF using a single thread and sequential writes.
We choose this scenario as the throughput remains unchanged regardless of whether block merging is enabled or not.
This ensures that the comparisons on the CPU overhead are relatively fair.
The X-axis of Figure~\ref{fig:motivation_merging} shows the number of 4~KB data blocks that can be potentially merged. 
We control this number via the \texttt{blk\_start\_plug} and \texttt{blk\_finish\_plug} function calls in the code.

We find that merging substantially reduces the CPU cycles of both the initiator and target, atop both flash and Optane SSD.
Merging decreases the number of requests and further NVMe-oF I/O commands, thereby reducing CPU cycles spent on the two-sided RDMA SEND operations. 
Although merging itself requires some investments in CPU cycles, it decreases CPU cycles to a greater extent, and thus the overall CPU overhead is improved.
We conclude that the I/O stack should merge ordered write requests to reduce CPU overhead for fast drives.
\section{\sysname Design}\label{sec:design}
Inspired by the studies in \S\ref{sec:motivation}, we introduce \sysname to preserve storage order  while taking advantage of fast NICs and SSDs.

\begin{figure}[t] 
	\centering 
	\includegraphics[width=\linewidth]{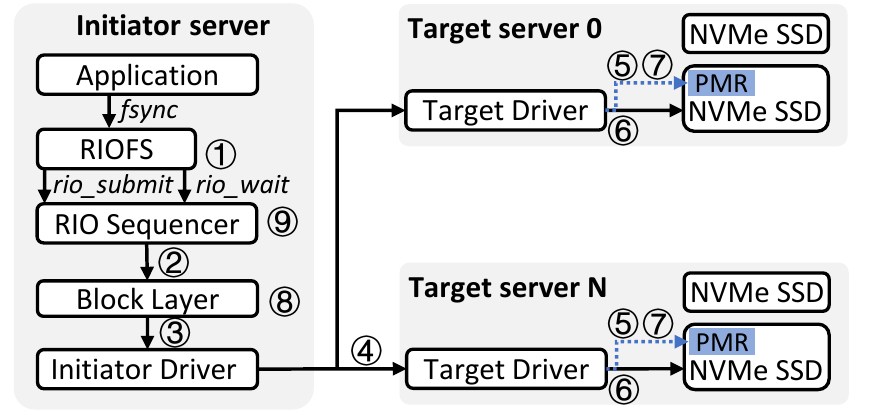} 
	\vspace{-15pt}
	\caption{\textbf{\sysname architecture.}} 
	\label{fig:design_overview} 
\end{figure}

\subsection{Overview}
Figure~\ref{fig:design_overview} shows the overview of \sysname, a networked storage stack that spans over one initiator server and multiple target servers.
Each target server consists of one or more NVMe SSDs, and requires at least one NVMe SSD with PMR support or a small region (2~MB) of byte-addressable persistent memory (e.g., Intel Optane main memory).
In the initiator server, we revise the entire storage stack including the file system, block layer and device driver, to let the ordering flow through the stack asynchronously.
Besides, a shim layer called \sysname sequencer between the file system and block device is introduced to package the original orderless block abstraction as an ordered one.
To benefit from \sysname, applications can invoke intact file system calls (e.g., \texttt{write}, \texttt{fsync}) on \fsname which leverages \sysname to accelerate the file system journaling.

The key design of \sysname is to control the order at the start and end of the lifetime of ordered write requests, while allowing some out-of-order execution in between.
Specifically, when ordered write requests are initiated by the file system or applications (\circled{1}), \sysname sequencer first generates a special \textit{ordering attribute} which is an identity of ordered request and used for reconstructing storage order, and then dispatches the requests to the block layer asynchronously (\circled{2}).
When ordered write requests are finished and returned to \sysname sequencer, \sysname completes the requests in order using the ordering attributes (\circled{9}), to handle the temporary out-of-order execution.
The file system and applications thus see the original ordered state.
Then, the intermediate execution (\circled{3}, \circled{4}, \circled{6} and \circled{8}) becomes almost asynchronous and concurrent, enabling more requests to be processed by each layer simultaneously (lessons 1 and 2 from \S\ref{sec:motivation}).

\sysname's approach, compared to existing designs, allows more outstanding requests to NICs and SSDs, thereby taking full advantage of the abundant concurrency of modern fast NICs and NVMe SSDs.
\sysname also enables easy scaling to more individual target servers, as there are no ordering constraints on the data transfer of ordered write requests (\circled{4}, \circled{6}).

The asynchronous execution makes the post-crash states of \sysname more uncertain, thus making crash consistency guarantees more challenging.
For example, after a server power outage, data blocks of the latter request are likely to be durable ahead of those of the former one, which violates the storage order.
\sysname addresses this issue with the persistent ordering attribute, which essentially logs the persistent state of data blocks of each ordered write request (\circled{5}, \circled{7}).
By scanning persistent ordering attributes, \sysname can speculate on possible post-crash states, and further recover data blocks to the latest and ordered state.
\sysname performs recovery I/Os in an asynchronous and concurrent fashion, thereby also fully utilizing the NICs and SSDs.

Making storage order asynchronous also brings opportunities.
The major opportunity is that, unlike traditional designs, consecutive ordered write requests of \sysname can be staged and merged to reduce CPU overhead.
For two consecutive ordered write requests, the classic NVMe over RDMA generates at least two NVMe-oF commands, which require at least four RDMA SEND operations. 
With RIO's I/O scheduler optimized for networked storage, the number of NVMe-oF commands and associated operations is halved. 
This further reduces the CPU cycles consumed at the device drivers.

In this section, we first present the organization of the ordering attribute (\S\ref{sec:design_attribute}) and the way of using it to enforce storage order (\S\ref{sec:design_parallel_persistence}). 
We next describe the crash recovery algorithm (\S\ref{sec:design_crash_recovery}) and show the I/O scheduling (\S\ref{sec:design_io_scheduler}).
We finally present the programming model (\S\ref{sec:programming_model}) and \fsname (\S\ref{sec:design_riofs}), and prove the correctness (\S\ref{sec:design_proof}), ending with discussion of support for multiple initiator servers (\S\ref{sec:design_discussion}).

\begin{figure}[t] 
	\centering 
	\includegraphics[width=\linewidth]{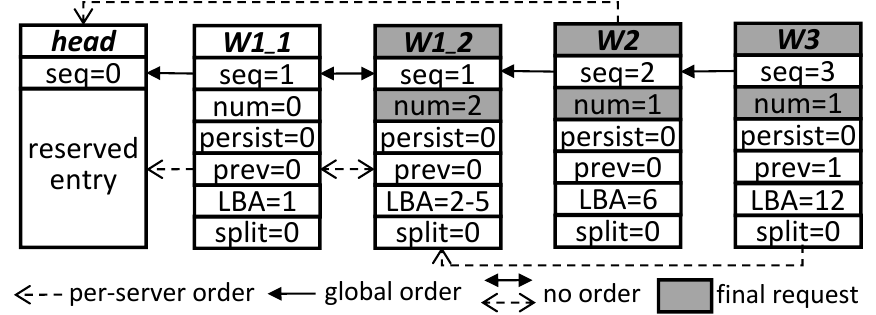} 
	\vspace{-10pt}
	\caption{\textbf{Ordering attributes.} \textit{Each big rectangle represents an ordering attribute of each ordered write request.}} 
	\vspace{-10pt}
	\label{fig:design_ordering_attributes} 
\end{figure}

\subsection{Ordering Attributes}\label{sec:design_attribute}
\noindent
\textbf{Definition.}
The ordering attribute is an ordered write request's logical identity that describes the group it belongs to, the previous request it follows, and whether associated data blocks are durable during  asynchronous execution.
As shown in Figure~\ref{fig:design_ordering_attributes}, ordering attributes essentially form two kinds of lists, one for global order and the other for per-target-server order.
The global order is the storage order for the entire cluster that consists of multiple target servers.
It is recorded by a sequence number (\texttt{seq}) widely used in the distributed environment.
The per-server order is the storage order for each target server.
It is achieved via the \texttt{prev} field, which points to its preceding request in the same target server.
An ordered write flow may contain several requests that can be freely reordered with each other, e.g., the journal description and journaled metadata.
\sysname groups this kind of request (e.g., \texttt{W1\_1} and \texttt{W1\_2}) using the same sequence number for each request and a counter (\texttt{num}) in the final request to record the number of requests in the group. 
\sysname guarantees storage order at the granularity of group.
By scanning and examining ordering attributes, the storage order can be built up and \sysname can regulate and recover the system to a correct state.

\noindent
\textbf{Creation.}
The ordering attribute is generated by the \sysname sequencer and embedded inside each request.
\sysname sequencer uses submission order from the file system as the storage order.
Specifically, a request that comes first is assigned a lower sequence number.
\sysname sequencer increases the sequence number and calculates the number of individual requests in its group when it encounters a special request that marks the end of a group of ordered write requests.
\sysname sequencer retains the most recent sequence number for each target server, to fill the \texttt{prev} field.
The \texttt{persist} field indicates whether the request is durable.
Its initial value is \texttt{0} and used for recovery.
The \texttt{LBA} field represents the logical block addresses of the request.
\sysname uses the \texttt{split} field to handle request splitting.

Though conceptually simple, ordering attributes are powerful across the I/O stack.
For example, they allow correct and parallel persistence and detection of out-of-order execution during recovery.
The block layer and device drivers leverage ordering attributes to perform merging and splitting, thus reducing CPU overhead and increasing I/O concurrency.

\subsection{Parallel and Correct Persistence}\label{sec:design_parallel_persistence}
Simply letting ordered write requests become orderless all the time leads to incorrect persistence and makes it difficult for the I/O stack to recover.
The key point here is that a consensus must be achieved on the persistence boundary between the software (target driver) and hardware (NVMe SSD), as the software and hardware have distinct contexts.
We introduce two techniques to achieve such a consensus.
First, the target driver submits the requests in per-server order (step \circled{6} of Figure~\ref{fig:design_overview}) to ensure correct persistence by the original durability interface (\S\ref{sec:design_in_order_submission}).
Second, since the NVMe SSD does not understand the ordering attribute, the ordering attribute needs to be documented so that it can be parsed by the software when recovery is performed (\S\ref{sec:persistent_order_attr}).

\subsubsection{In-Order Submission}\label{sec:design_in_order_submission}
\sysname keeps the multi-queue design from the RDMA and NVMe stack. 
An ordered write request can be distributed to any hardware queue and an RDMA NIC is likely to reorder requests among multiple queues (step \circled{4}).
Assume \texttt{W3} of Figure~\ref{fig:design_ordering_attributes} arrives at the target server earlier than \texttt{W1\_2} and requires instant data persistence to flush the SSD, i.e., requests before \texttt{W3} must be stored in persistent media rather than the SSD's volatile write buffer.
If the target driver directly submits \texttt{W3}, the SSD only makes \texttt{W1\_1} and \texttt{W3} durable by a \texttt{FLUSH} command, ignoring \texttt{W1\_2} and violating the original durability semantics.

To address this issue, the target driver of \sysname submits ordered write requests to the SSD in the per-server order. 
In the aforementioned example, the target will not submit \texttt{W3} until all \texttt{W1\_1} and \texttt{W1\_2} are dispatched to the SSD.

The in-order submission mechanism allows each target server to transfer data blocks and flush SSDs concurrently.
\sysname does not use the global order of submission for avoiding coordination among servers.
If the global order is used, the target server that has \texttt{W3} must wait for the target server that has \texttt{W2}.
This not only incurs extra network traffic but also introduces synchronization overhead, lowering the concurrency.
As we show later, \sysname can recover the out-of-order persistence over multiple servers in case of a crash.

Maintaining the per-server submission order potentially introduces synchronization overhead, as a later request is blocked until its preceding requests reach the server.
Fortunately, this overhead is negligible due to the massive concurrency of NICs. 
The concurrency of NICs is usually larger than SSDs installed on the same server, and therefore the NIC can post storage I/Os to the target driver almost at the same time.
We further use the in-order delivery property of RDMA to remove this overhead (\S\ref{sec:design_io_scheduler}).

\subsubsection{Persistent Ordering Attributes}\label{sec:persistent_order_attr}
\sysname makes ordering attributes persistent so as to reconstruct per-server ordering lists for further recovery.
The key idea is to record the persistence state of data blocks in the \texttt{persist} field of the associated ordering attribute.
Before submitting an ordered request to SSD, \sysname persists the ordering attribute (step \circled{5}), which logs the storage order but indicates that data blocks are still in-progress and thus non-persistent.
When data blocks become persistent, \sysname sets the \texttt{persist} field to \texttt{1} (step \circled{7}).
Specifically, for SSDs with power loss protection (PLP), e.g., non-volatile write cache, \sysname toggles the \texttt{persist} field when a completion response is reported via the interrupt handler, since data blocks become durable when they reach the SSD and the \texttt{FLUSH} command is usually ignored by the block layer.
For SSDs without PLP, the \texttt{persist} field is set only after the request with a \texttt{FLUSH} command is completed.
Only one \texttt{persist} field whose request has a \texttt{FLUSH} command is toggled to indicate that data blocks of all preceding write requests become durable.

\sysname stores the persistent ordering attributes to the persistent memory region (PMR) of NVMe SSDs.
In particular, \sysname organizes the PMR as a circular log and employs two in-memory pointers, the \texttt{head} and \texttt{tail} pointers, to efficiently manage the circular log.
\sysname appends the newly arrived ordering attributes to the log tail by increasing the \texttt{tail} pointer.
Once the completion response of the ordered write request is returned to the application (indicating that the storage order is satisfied), associated ordering attributes become invalid and \sysname recycles space by moving the \texttt{head} pointer.

In \sysname, storing the small-sized ordering attributes by CPU-initiated MMIOs (usually less than 1 $\mu$s) is significantly faster than persisting data blocks (usually more than 10 $\mu$s).
Moreover, each target server persists ordering attributes in parallel without any coordination.
Hence, storing ordering attributes does not introduce much overhead to the overall I/O path.

Persistent ordering attributes are used to rebuild per-server ordering lists in parallel.
Specifically, each server validates each ordering attribute by checking the \texttt{persist} field.
For SSDs with PLP, an ordering attribute is valid if and only if its and its preceding attribute's \texttt{persist} fields are all set to \texttt{1}.
For SSDs without PLP, an ordering attribute is valid when the \texttt{persist} field of a latter ordering attribute that belongs to a \texttt{FLUSH} command is set to \texttt{1}.
By scanning ordering attributes, valid per-server storage order can be reconstructed.
Other invalid ordering attributes are dropped, as the storage order among the ordered write requests is uncertain.
With parallel persistence and validation of ordering attributes, \sysname processes ordered write requests with high concurrency.

\begin{figure}[t] 
	\centering 
	\includegraphics[width=\linewidth]{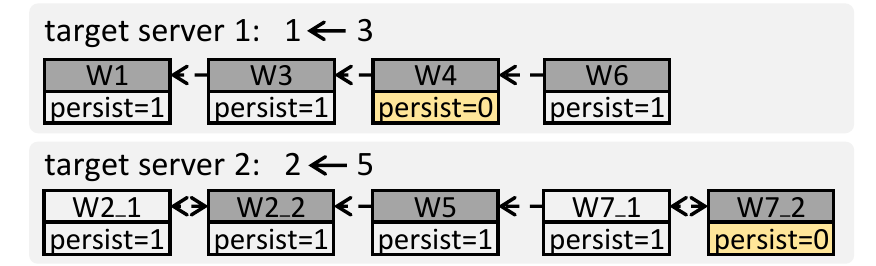} 
	\vspace{-15pt}
	\caption{\textbf{A recovery example.} \textit{Other fields of the ordering attributes are omitted to facilitate the description.}} 
	\vspace{-10pt}
	\label{fig:design_recovery} 
\end{figure}

\subsection{Crash Recovery and Consistency}\label{sec:design_crash_recovery}
The storage system must be able to recover to a consistent state in the face of a sudden server crash (e.g., a power outage).
The main idea is to leverage reconstructed per-server ordering lists from target servers (described in \S\ref{sec:persistent_order_attr}) to detect out-of-order execution, so as to perform crash recovery.
In this section, we first present RIO's crash recovery and consistency on systems that update out-of-place (e.g., log structure), and then show how RIO handles in-place updates.

\subsubsection{Out-of-Place Updates.}
We use Figure~\ref{fig:design_recovery} to elaborate on \sysname's recovery.
In this figure, we assume that ordered write requests of \sysname always update out-of-place so that there always exists one copy of old data blocks.
As both the initiator and target servers can fail, we describe \sysname's recovery strategy in these two types of crash scenarios.

\noindent
\textbf{Initiator recovery.}
After restarting and reconnecting to all target servers, the initiator server collects per-server ordering lists (1$\leftarrow$3, 2$\leftarrow$5) from each target server.
Next, the initiator rebuilds a global ordering list (1$\leftarrow$2$\leftarrow$3) by merging per-server ordering lists.
For example, \texttt{W5} is dropped since \texttt{W4} is not durable.
Then, the global ordering list is sent back to each target server to roll back.
Data blocks that are not within the global ordering list (\texttt{W4}, \texttt{W5}, \texttt{W6} and \texttt{W7}) are erased.

\noindent
\textbf{Target recovery.}
\noindent
When a target server crashes, the initiator server tries to reconnect the target server.
Once connected again, similar to the initiator recovery, the initiator server firstly rebuilds the global ordering list.
The difference is that merging does not drop ordering attributes of alive target servers.
Instead, the initiator server tries to repair the broken list by replaying non-persistent requests on failed targets.
For example, assume target server 1 fails while server 2 is still alive.
The initiator re-sends \texttt{W4} to target 1 until a successful completion response is received.
Replaying is idempotent and thus does not introduce inconsistency.

The version \textbf{consistency} requires that metadata is consistent and the versions of data and metadata match with each other.
Most mechanisms that support version consistency (e.g., the data journaling of Ext4, the checkpointing of F2FS~\cite{F2FS_fast15}) use the storage order to keep the versions of data and metadata the same and update data and metadata blocks out-of-place for crash recovery. 
\sysname provides storage order and is thus capable of offering version consistency when the data and metadata blocks are updated out-of-place.

\subsubsection{In-Place Updates (IPUs)}
Crash-consistent storage systems atop commodity SSDs that do not have an atomic update interface usually update metadata out-of-place for system integrity.
For user data, IPUs can be classified into two categories: normal IPUs that overwrite an existing file, and block reuse where data blocks of a file are re-assigned to another file.
\sysname is unaware of IPUs and upper layer systems (e.g., file systems) need to explicitly pass labels.
\sysname distinguishes IPUs from out-of-place updates by a special field (\textit{ipu}) in the ordering attribute and handles them differently.

The target recovery for IPUs is the same as out-of-place updates. 
However, the initiator recovery is different: 
\sysname does not perform roll-back but leaves the recovery strategy to upper layer systems (e.g., file systems).
The file system can thus retrieve the global ordering list from \sysname to achieve a certain level of consistency by customizing recovery logic.

This design of \sysname leaves flexible consistency guarantees to upper layer systems, as handling IPUs is tricky.
For example, data consistency (e.g., ordered mode of Ext4) requires that metadata is consistent and data is persisted before metadata.
Ext4 achieves data consistency by updating data in-place before writing metadata to the journal area.
As the global ordering list provides the persistence order, a file system built atop \sysname can also achieve data consistency by erasing the metadata blocks that are persisted before IPU data blocks during recovery.

For block reuse, upper layer systems can not directly use \sysname since the newer data can not be durable before the data block is freed from the prior owner (otherwise the prior owner can see the data of the new owner).
For example, a thread issues a request to change the ownership of a file, followed by issuing another request to write data to that file.
The file system uses \sysname to order the two requests.
If a crash happens when data blocks of the later request are durable while the former is not durable, the file system will fail to recover to a consistent state.
If the file system erases the data blocks of the later requests, the data content of the prior owner is lost.
If the file system leaves the data blocks of the later requests untouched, the prior owner can see the data content of the new owner, which results in security issues.

Upper layer systems (e.g., file systems) need to regress to the data journaling or use classic synchronous \texttt{FLUSH} for block reuse.
For example, a file system can write the data blocks to the journal area to keep an old copy, so that the file system can roll back to the old copy during recovery if the metadata block that describes the file ownership is not durable.
A file system can also issue a \texttt{FLUSH} command and wait for its completion to ensure that the ownership changes before the new data are written to the file.

\begin{figure}[t] 
	\centering 
	\includegraphics[width=\linewidth]{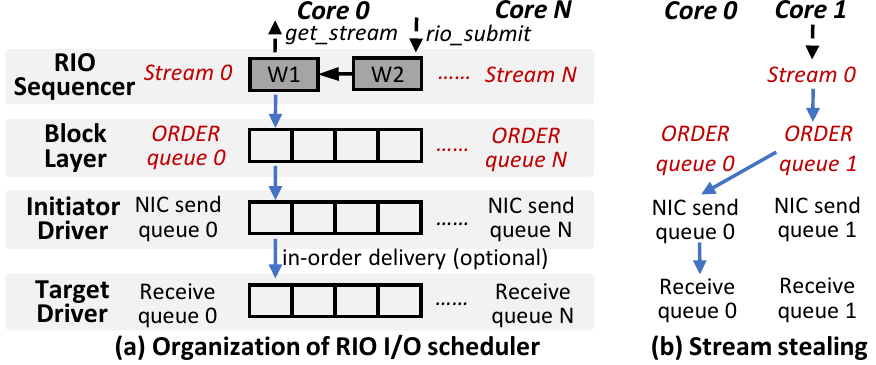} 
	\vspace{-15pt}
	\caption{\textbf{\sysname I/O scheduler.}
	\textit{(a) The organization of \sysname I/O scheduler. (b) The way of \sysname for handling thread migration.}} 
	\vspace{-10pt}
	\label{fig:design_io_scheduler} 
\end{figure}

\subsection{\sysname I/O Scheduler}\label{sec:design_io_scheduler}
\sysname I/O scheduler is designed to exploit the asynchronous execution of ordered write requests and reduce the CPU overhead of the networked storage stack.

Figure~\ref{fig:design_io_scheduler}(a) presents the organization of \sysname I/O scheduler.
We use the \textit{stream} notion for multicore scalability.
The stream represents a sequence of ordered write requests. 
Across streams, there are no ordering restrictions, i.e., each stream has independent global order.
The number of streams is configurable and by default equals the number of cores.
Each CPU core can get arbitrary streams but has a dedicated stream in the common case, i.e., core 0 uses stream 0.
\sysname I/O scheduler has three main design principles.

\noindent
\textbf{Principle 1:} \sysname uses dedicated software queues (\textit{ORDER queue}) to schedule ordered write requests. 
Such separation of ordered requests from orderless ones reduces the latency of ordered write requests which are usually on the critical path (e.g., \texttt{fsync}), and simplifies the scheduling algorithms.

\noindent
\textbf{Principle 2:} \sysname dispatches requests of a stream to the same NIC send queue, to exploit the in-order delivery property of the network protocol, thereby reducing the overhead of in-order submission of the target driver (\S\ref{sec:design_in_order_submission}).
For example, the block layer dispatches requests from stream 0 to NIC queue 0.
As the reliable connected (RC) transport of RDMA preserves the delivery order of RDMA SEND operations for each queue, aligning the stream to a NIC queue reduces out-of-order deliveries over the network. 
Each socket of the TCP stack has similar in-order delivery property.
Thus, this principle can be applied to TCP networks.

\noindent
\textbf{Principle 3:} The merging and splitting of \sysname may enhance (but must not sacrifice) the original ordering guarantees.
For example, merging adjacent data blocks of continuous ordered write requests may remove the storage barrier.
The merged request however should be atomic since atomicity is a stronger consistency property than storage order.

\begin{figure}[t] 
	\centering 
	\includegraphics[width=\linewidth]{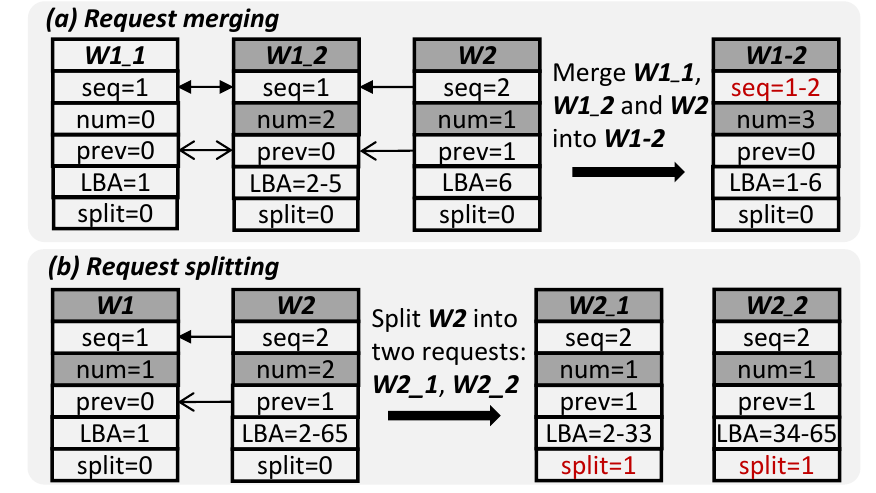} 
	\vspace{-15pt}
	\caption{\textbf{Request merging and splitting in \sysname.}
	\textit{The `persist' field is initialized to 0 and is omitted for simplicity.}} 
	\vspace{-15pt}
	\label{fig:design_blk_merge} 
\end{figure}

Whenever possible, \sysname merges consecutive ordered write requests into a single large request.
This reduces the number of NVMe-oF I/O commands and further CPU cycles on the storage protocol (lesson 3 from \S\ref{sec:analysis_lessons}).
We use examples from Figure~\ref{fig:design_blk_merge} to illustrate principle 3 at length.
By default, \sysname does not reorder requests in the \textit{ORDER} queue.
However, reordering is allowed in \sysname for fairness and rate limit but this is beyond the scope of this paper.

\noindent
\textbf{Request merging.} There are three requirements for request merging.
First, merging is performed within a sole stream.
Second, sequence numbers of requests must be continuous in order not to sacrifice the original storage order.
Third, logical block addresses (LBAs) of requests must be non-overlapping and consecutive.
Figure~\ref{fig:design_blk_merge}(a) shows three requests that meet these three requirements.
The block layer merges them in the \textit{ORDER} queue and compacts their ordering attributes into a single one.
If merging spans multiple groups, the sequence number of the merged request becomes a range.
As the three requests share a sole ordering attribute, they are considered as a whole during recovery and thus become atomic.
In particular, the \texttt{persist} field will only be toggled if all three requests are durable.
Otherwise, all three requests are discarded or replayed during recovery.

\noindent
\textbf{Request splitting.} 
The block layer of \sysname splits a request to meet the hardware limitation (e.g., the transfer size of a single request) and software configuration (e.g., the stripe size of a logical volume).
For example, the maximum transfer size of a single request of an Intel 905P SSD is 128~KB.
An RDMA NIC has a similar constraint for a single work request.
\sysname divides the larger ordered write request into smaller ones and tags the divided requests with a special \texttt{split} flag.
During recovery, divided requests are merged back into the original request to validate the global order.
For example in Figure~\ref{fig:design_blk_merge}(b), \texttt{W2} is divided and scattered to two servers. 
During crash recovery, ordering attributes of \texttt{W2\_1} and \texttt{W2\_2} are sent back to the initiator to decide the global order.

A merged request can not be split, and vice versa. 

A process can be migrated from one core to another due to the CPU scheduling, which leads to stream stealing (Figure~\ref{fig:design_io_scheduler}(b)) and may complicate the I/O scheduler and make the stream notion difficult to use.
To handle this case, \sysname allows stream stealing and affiliates the stream to NIC send queue, which always forwards requests of a stream to the same NIC queue regardless of process's migration.
Similar to the orderless requests, the pending requests in the \textit{ORDER} queue are flushed to the initiator driver before the process is migrated.

\subsection{Programming Model}\label{sec:programming_model}
\sysname provides an ordered block device abstraction and asynchronous I/O interfaces to file systems and applications.
Specifically, the \texttt{rio\_setup} function specifies the number of streams, ideally to the number of independent transactions allowed in the applications or file systems (e.g., the number of cores) to maximize concurrency.
The \texttt{rio\_setup} function also associates the networked storage devices (e.g., a sole SSD, a logical volume or RAID) with the streams.
The \texttt{rio\_submit} function encapsulates the original block I/O submission function (\texttt{submit\_bio}).
It requires a stream ID and a flag to explicitly delimit the end of a group.
\sysname treats the submission order from file systems (or applications) as the global order and automatically manages the per-server and global order for each stream.
File systems and applications only need to manage streams and decide the submission order. 
The \texttt{rio\_submit} function dispatches requests to the target with ordering guarantees.
To guarantee durability, file systems and applications need to embed a \texttt{FLUSH} command in the final request and use \texttt{rio\_wait} to poll for the completion of the final request.
The users can continuously push multiple asynchronous and ordered requests to SSDs via \texttt{rio\_submit} and use \texttt{rio\_wait} for durability, thereby achieving high concurrency.

\begin{figure}[t] 
	\centering 
	\includegraphics[width=\linewidth]{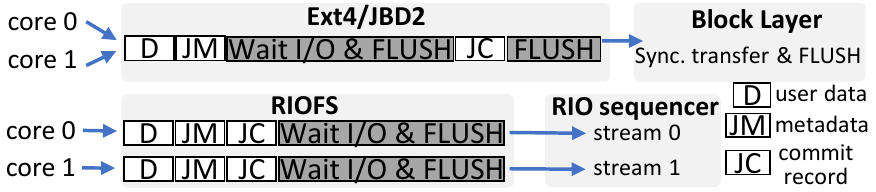} 
	\vspace{-15pt}
	\caption{\textbf{\fsname atop \sysname.}
	\textit{\fsname uses the \sysname sequencer to perform file system journaling.}} 
	\vspace{-15pt}
	\label{fig:design_riofs} 
\end{figure}

A typical use case that relies heavily on the storage order is file system journaling (or write-ahead logging).
File systems can associate each independent on-disk journaling with each stream and use \texttt{rio\_submit} to dispatch journaling I/Os.
We show a detailed file system design atop \sysname in \S\ref{sec:design_riofs}.
Applications that are built atop the block device (e.g., BlueStore~\cite{bluestore}) can also use \sysname to accelerate on-disk transactions.
For example, applications can replace the asynchronous I/O interfaces (e.g., libaio~\cite{libaio}) with \texttt{librio}, which is a wrapper of the in-kernel interfaces such as \texttt{rio\_submit}.

\subsection{\fsname: A File System atop \sysname}\label{sec:design_riofs}
In this section, we introduce \fsname to alleviate the performance bottleneck from the file system and evaluate unmodified applications atop \sysname.
We develop \fsname based on Ext4 and use two main techniques (Figure~\ref{fig:design_riofs}).
First, \fsname replaces all synchronous transfer and \texttt{FLUSH} commands that are used for storage order with the stream interfaces (e.g., \texttt{rio\_submit}).
This parallels ordered write requests of a single transaction.
Second, \fsname employs a per-core journaling design from iJournaling~\cite{iJournaling-atc17} to increase the multicore scalability.
Specifically, each core has a private journal area and performs journaling almost independently.
When \texttt{fsync} is called, the thread dispatches only the corresponding file-level transaction to its dedicated stream.
To handle journal conflicts (e.g., multiple transactions on an inode), \fsname compares the global transaction IDs (sub-transaction IDs) and applies the latest transaction during checkpoint or recovery, following the iJournaling's design.
The only difference between iJournaling and \fsname is the method of storage ordering guarantees. iJournaling uses synchronous transfer and \texttt{FLUSH} commands while \fsname uses the \sysname sequencer.

\fsname uses \sysname to handle normal IPUs, but regresses to the classic approach (i.e., synchronous \texttt{FLUSH}) to handle block reuse. 
Performing \texttt{FLUSH} for block reuse will not harm the average throughput unless the file system is nearly 100\% full. 
In normal situations, \fsname first find free data blocks that are not referenced by any files without \texttt{FLUSH} commands.

\subsection{Proof of \sysname's and \fsname's Correctness}\label{sec:design_proof}
This section proves the correctness of \sysname approach to storage order.
We refer to data blocks of an ordered write request as ${D_n}$.
The \texttt{n} represents the global order, i.e., the \texttt{seq} value.
We refer to the associated ordering attribute of the request as ${O_n}$.
When the \texttt{persist} value is \texttt{0}, i.e., associated data blocks are not durable, the ordering attribute is $\bar{O_n}$.
We use the term $\leftarrow$ to describe the ``persist-before'' relationship;
${D_{n-1}}$ $\leftarrow$ ${D_n}$ means ${D_{n-1}}$ must be durable prior to ${D_{n}}$.
We thus have $\bar{O_n}$ $\leftarrow$ ${D_n}$ $\leftarrow$ ${O_n}$ (steps \circled{5}, \circled{6} and \circled{7} of Figure~\ref{fig:design_overview} and \S\ref{sec:design_parallel_persistence}).

To prove the correctness of \sysname, we only need to prove that the post-crash state of \sysname is valid, as the in-order completion mechanism of \sysname guarantees an ordered state during normal execution.
Assume there are \texttt{N} ordered write requests.
Then, there are \texttt{N+1} valid post-crash states, $\emptyset$, $D_1$, ... , $D_1$ $\leftarrow$ $D_2$ $\leftarrow$ ... $D_k$, ... , $D_1$ $\leftarrow$ $D_2$ $\leftarrow$ ... $D_n$.
All states preserve prefix semantics. 
Other states, e.g., $D_{k+1}$ $\leftarrow$ $D_{k}$, are invalid.

We consider the basic case with no merging and splitting first. 
During crash recovery, \sysname first scans ordering attributes from ${O_1}$ to ${O_n}$.
Suppose it finds that the first non-durable ordering attribute is $\bar{O_k}$.
In other words, preceding ordering attributes are ${O_1}$, ${O_2}$, ... ${O_{k-1}}$. As ${D_n}$ $\leftarrow$ ${O_n}$, data blocks of the former \texttt{k-1} requests are durable, i.e., ${D_1}$ $\leftarrow$ ${D_2}$ $\leftarrow$ ... ${D_{k-1}}$.
This is a valid state that obeys the storage order.
Due to the asynchronous execution, data blocks of a request later than the \texttt{k}th can be durable. 
Suppose this request is the \texttt{m}th, and thus we have $O_m$, $m$ > $k$ and $D_m$ $\leftarrow$ $D_k$ which disobeys the storage order. As $\bar{O_m}$ $\leftarrow$ $D_m$ and $\bar{O_m}$ already records the locations of $D_m$, \sysname performs recovery algorithm to discard $D_m$ or replay $D_k$ ... $D_{m-1}$.
As a result, the post-crash state remains ${D_1}$ $\leftarrow$ ${D_2}$ $\leftarrow$ ... ${D_{k-1}}$ by discarding, or changes to ${D_1}$ $\leftarrow$ ${D_2}$ $\leftarrow$ ... ${D_m}$ by replaying (\S\ref{sec:design_crash_recovery}).
Both post-crash states are valid and therefore \sysname preserves the storage order.

Recall that \sysname can only merge data blocks of consecutive requests (principle 3 from \S\ref{sec:design_io_scheduler}). 
Assume \sysname merges $D_k$, $D_{k+1}$, ... $D_m$ into $D^m_k$, where $D^m_k$ indicates data blocks from request \texttt{k} to \texttt{m}.
Thus, associated ordering attributes are also merged into $O^m_k$.
During crash recovery, $O^m_k$ is considered as one sole ordering attribute.
Then, the proof returns to the aforementioned basic case.
The only difference is that the consistency guarantee among $D_k$ to $D_m$ is enhanced to atomicity.
In particular, there are \texttt{m-k+2} valid post-crash states without merging, $\emptyset$, $D_k$, $D_k$ $\leftarrow$ $D_{k+1}$, ... , $D_k$ $\leftarrow$ $D_{k+1}$ $\leftarrow$ ... $D_m$. With \sysname's merging, the number of post-crash states is reduced to 2.
The states are $\emptyset$ or $D_k$ $\leftarrow$ $D_{k+1}$ $\leftarrow$ ... $D_m$, representing the ``nothing'' or ``all'' states of the atomicity guarantee, respectively.

Recall that \sysname merges the divided requests back to the original request when performing recovery. As a result, the proof also returns to the basic case if a request is split.

The correctness of \fsname depends on \sysname and iJournaling.
For storage order, \fsname replaces the \texttt{FLUSH} commands with \sysname's ordering primitive to connect the file system (iJournaling) to the block layer.
Since \sysname guarantees storage order and iJournaling is correct, \fsname is correct.

\subsection{Discussion}\label{sec:design_discussion}
\sysname assumes a sole initiator server accessing an array of SSDs. 
Distributed concurrency control over multiple initiator servers is orthogonal to this paper, and will not be the slower part that affects the overall performance as it is performed mostly in memory which is significantly faster than remote storage access (1M ops/s). 
For example, the throughput of allocating a sequencer number approaches 100M ops/sec~\cite{rdmabench-atc16}.
\sysname's architecture (Figure~\ref{fig:design_overview}) can be extended to support multiple initiator servers, by extending \sysname sequencer and \fsname to distributed services.
We leave this for future work.
\section{\sysname Implementation}\label{sec:imple}

\begin{table}[t]
	\centering
    \small
    \begin{threeparttable}[t]
	\begin{tabular}{l|c|c}
		\hline
		\textbf{Dword:bits} & \textbf{NVMe-oF} & \textbf{\sysname NVMe-oF} \\ \hline \hline
		00:10-13 & reserved & \makecell[c]{\sysname op code, e.g., submit} \\ 
        02:00-31 & reserved & start sequence (\texttt{seq})\\
        03:00-31 & reserved & end sequence (\texttt{seq})\\ 
        04:00-31 & metadata\tnote{*} & previous group (\texttt{prev}) \\ 
        05:00-15 & metadata\tnote{*} & number of requests (\texttt{num}) \\ 
        05:16-31 & metadata\tnote{*} & stream ID \\ 
        12:16-19 & reserved & special flags, e.g., boundary \\ \hline
	\end{tabular}
    \begin{tablenotes}
		\item[*] The metadata field of NVMe-oF is reserved.
	\end{tablenotes}
	\vspace{8pt}
	\caption{\textbf{\sysname NVMe-oF commands atop 1.4 spec.}
	\textit{\sysname uses the reserved fields of the NVMe-oF I/O commands to transfer ordering attributes over the network.}}
	\label{tab:rio_command_format}
	\vspace{-15pt}
    \end{threeparttable}
\end{table}

We implement \sysname based on NVMe over RDMA driver from Mellanox~\cite{mellanox-nof-driver} and Linux kernel 4.18.20 on real hardware.

A critical issue is to pass ordering attributes across the I/O stack.
To distinguish ordered requests from the orderless, \sysname uses two special flags to represent normal and final ordered write requests.
\sysname sequencer uses the private field (\texttt{bi\_private}) of the original block I/O data structure (\texttt{bio}) to store ordering attributes and original private information together.
The block layer can therefore get ordering attributes from the \texttt{bio} struct and perform scheduling.
The initiator driver passes ordering attributes using reserved fields of the NVMe-oF write command.
The detailed command format is presented in Table~\ref{tab:rio_command_format}.
The target driver persists ordering attributes in PMR via persistent MMIO write (i.e., an MMIO read after an MMIO write).
As commercial SSDs used in the experiments do not support the relatively new PMR feature (released in NVMe spec 1.4 circa June 2019), we use 2~MB in-SSD capacitor-backed DRAM to serve as the PMR region for commercial SSDs, which is the same as in \horae~\cite{horae-osdi20} and ccNVMe~\cite{ccNVMe-sosp21}.
Specifically, \sysname remaps the non-volatile DRAM of an OC-SSD by the PCIe Base Address Register technique, which is mature and widely used by SSDs and NICs (e.g., doorbells).
The ordering attributes are written to PMR and data blocks are sent to commercial SSDs.


\begin{figure*}[t] 
	\centering 
	\includegraphics[width=\linewidth]{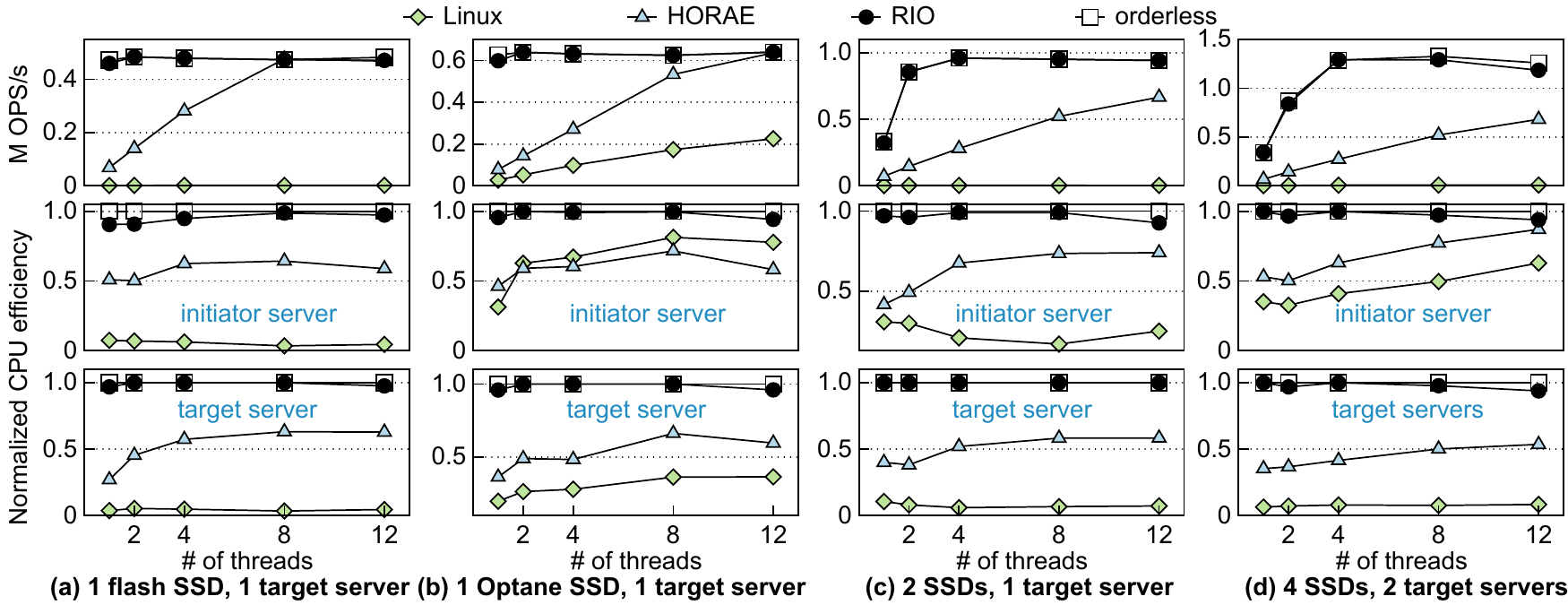} 
	\vspace{-18pt}
	\caption{\textbf{Block device performance.}
    \textit{4~KB random ordered write. CPU efficiency is normalized to the orderless.}} 
	\label{fig:eval_bdev_thread} 
\end{figure*}

\section{Evaluation}\label{sec:evaluation}
In this section, we first describe the setups of the test environment (\S\ref{sec:eval_setup}).
Next, we evaluate the performance of \sysname atop both flash and Optane SSDs, as well as atop both storage arrays in a single target server and across multiple target servers (\S\ref{sec:eval_block_device}).
Then, we examine the performance of \fsname through microbenchmarks (\S\ref{sec:eval_fs}) and applications (\S\ref{sec:eval_app}).
We finally study the recovery overhead of \sysname (\S\ref{sec:eval_recovery}).


\subsection{Experimental Setup}\label{sec:eval_setup}
\noindent
\textbf{Hardware.}
We conduct all experiments in three physical servers.
One is the initiator and the other two are target servers.
Each server has 2 Intel Xeon Gold 5220 CPUs, and each CPU has 18 cores and runs at 2.20~GHz.
We test three kinds of SSDs.
Target server 1 has one Samsung PM981 flash and one Intel 905P Optane SSDs.
Target server 2 has one Samsung PM981 flash and one Intel P4800X Optane SSDs.
We use 2~MB PMR for each target server, the same as \horae.
It costs around 0.6~$\mu$s to persist a 32~B ordering attribute to PMR.
The servers connect to each other with one 200~Gbps Mellanox ConnectX-6 RDMA NIC.

\noindent
\textbf{Compared systems.}
To evaluate the performance of ordered write requests atop block devices, we compare \sysname against the Linux NVMe over RDMA.
We also extend \horae to the same NVMe over RDMA stack.
Specifically, the control path is built atop the initiator driver, and uses two-sided RDMA SEND operations to transfer the ordering metadata.
When ordering metadata arrives at the target server, the target driver forwards it to PMR by a persistent MMIO write.
For file system and application performance, we compare \fsname against Ext4 and \horaefs.
To ensure the fairness of comparisons, we also adopt iJournaling's design in \horaefs, the same as \fsname (\S\ref{sec:design_riofs}). 
All three file systems are based on the same codebase of Ext4 and the same OS, and use metadata journaling and 1~GB journal space in total.
Both \fsname and \horaefs allocate 24 streams during the evaluation, which are enough to drive all 4 SSDs on the targets in most cases.

\noindent
\textbf{CPU efficiency.}
We use CPU efficiency to quantify the ability of storage systems to drive the I/O devices using a single unit of CPU cycles. 
Specifically, CPU efficiency is consistent with the write requests each CPU cycle can serve, i.e., throughput  $\div$ CPU utilization.
The CPU utilization is collected by \texttt{top} command.

\subsection{Block Device Performance}\label{sec:eval_block_device}
\begin{figure}[t] 
	\centering 
	\includegraphics[width=\linewidth]{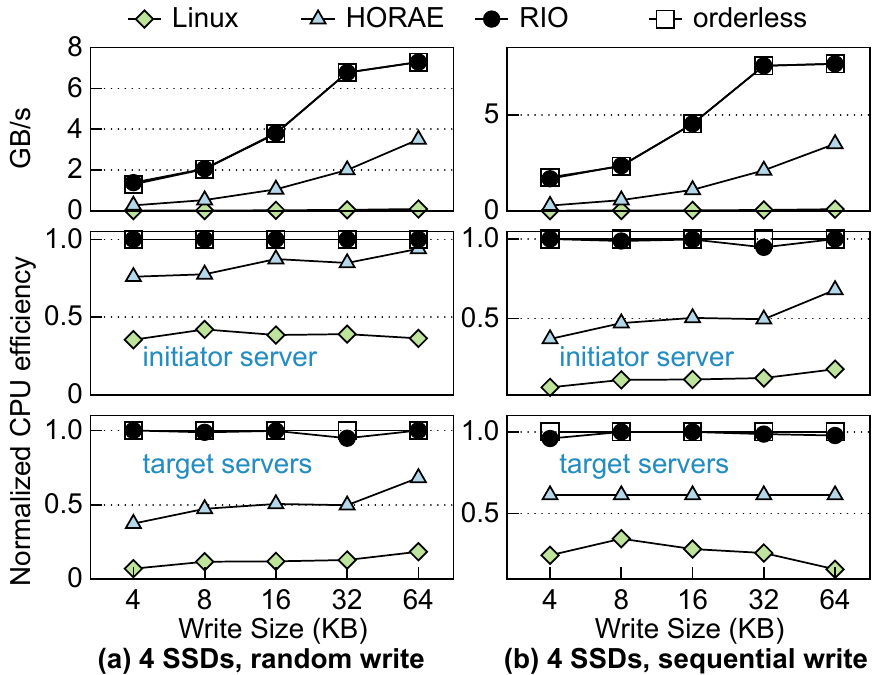} 
	\vspace{-18pt}
	\caption{\textbf{Performance with varying write sizes.}
	\textit{CPU efficiency is normalized to the orderless.}} 
	\label{fig:eval_bdev_write_size} 
\end{figure}

We examine the performance of block devices with different ways of storage ordering guarantees: Linux NVMe over RDMA, \horae and \sysname.
We also measure the performance of orderless write requests, to quantify how far we are from the optimal performance of ordered write requests. 
We conduct the experiments by varying the number of threads, the write size and the batch size of each group while keeping other parameters constant.
We collect the throughput and CPU utilization.
Figures~\ref{fig:eval_bdev_thread},~\ref{fig:eval_bdev_write_size} and~\ref{fig:eval_bdev_batch_size} illustrate the results across a variety of configurations.
We get three major findings: (1) \sysname achieves significantly larger I/O throughput with higher CPU efficiency against its peers; 
(2) the throughput and CPU efficiency of \sysname come close to the orderless;
(3) the \sysname I/O scheduler greatly increases the CPU efficiency and further boosts the throughput.
We next describe each figure in detail.

\subsubsection{Multicore performance}\label{sec:eval_block_device_thread}
Figures~\ref{fig:eval_bdev_thread}(a)-(d) plot the throughput and CPU efficiency with different numbers of threads.
Each thread submits random ordered write requests to an individual stream.

In the flash SSD (Figure~\ref{fig:eval_bdev_thread}(a)), \sysname achieves two orders of magnitude higher throughput than Linux NVMe-oF and outperforms \horae by 2.8$\times$ on average.
\sysname offers higher throughput with 18.0$\times$ and 1.7$\times$ CPU efficiency in the initiator server, and 22.7$\times$ and 2.1$\times$ CPU efficiency in the target server on average compared to Linux and \horae, respectively.
\sysname prevails over its peers for two reasons.
First, \sysname removes the prohibitive \texttt{FLUSH} command, which is usually a device-wide operation of the SSDs that do not have power loss protection (PLP) (e.g., capacitors for data since the last \texttt{FLUSH}).
The tested flash SSD does not have PLP, and whenever it receives a \texttt{FLUSH} command, it instantly drains off the data in the volatile buffer to the persistent flash memory, which neutralizes the multicore concurrency from the host and multi-chip concurrency from the flash when little data is buffered.
Second, \sysname makes the ordered I/O path mostly asynchronous and thus fully exploits the bandwidth of the NIC and SSD.
We observe that the throughput of \horae is significantly lower than \sysname when the count of threads is small, due to synchronous execution of the control path. 
The control path also decreases the CPU efficiency as the control path incurs additional network traffic (e.g., RDMA SEND) and therefore demands more CPU cycles.
\sysname reuses the functions of request merging from the orderless.
By comparing the CPU efficiency of these two systems, we find the additional logic that the \sysname I/O schedule adds (e.g., comparing the ordering attributes) does not introduce much overhead to both the CPU and I/O performance.

In the Optane SSD (Figure~\ref{fig:eval_bdev_thread}(b)), \sysname exhibits 9.4$\times$ and 3.3$\times$ throughput on average against Linux and \horae.
\sysname delivers 1.7$\times$ and 1.7$\times$ CPU efficiency in the initiator server, and 3.5$\times$ and 2.0$\times$ CPU efficiency in the target server compared to Linux and \horae, respectively.
Here, the Optane SSD has PLP and thus the \texttt{FLUSH} does not influence the throughput significantly.
Yet, the synchronous execution is still a dominant factor affecting the overall throughput and CPU efficiency.
By dramatically reducing the proportion of synchronous execution, \sysname shows similar throughput and CPU efficiency against the orderless.

We extend the experiments to multiple SSDs (Figure~\ref{fig:eval_bdev_thread}(c)) and two target servers (Figure~\ref{fig:eval_bdev_thread}(d)). 
The SSDs are organized as a single logical volume and the tested systems distribute 4~KB data blocks to individual physical SSDs in a round-robin fashion.
Linux can not dispatch the following ordered write request to other SSDs until the previous one finishes.
\horae can not dispatch ordered write requests to SSDs in parallel before the control path finishes.
Unlike Linux and \horae, \sysname can distribute ordered write requests to SSDs concurrently and hence shows high CPU efficiency and I/O throughput.
As we add more SSDs, a single thread can not fully utilize the overall bandwidth, even for the orderless.
In this case, CPU efficiency becomes a dominant factor that affects the I/O throughput.
\sysname fully drives the SSDs with 4 threads due to high CPU efficiency.

\subsubsection{Performance with varying write sizes}\label{sec:eval_block_device_write_size}
Figure~\ref{fig:eval_bdev_write_size} presents the throughput and CPU efficiency with varying write sizes.
Only one thread is launched to perform sequential and random ordered writes.

We get similar results as in \S\ref{sec:eval_block_device_thread}.
Specifically, \sysname outperforms Linux and \horae by up to two orders of magnitude and 6.1$\times$ respectively, with a more efficient use of CPU cycles.
The key takeaway here is that asynchronous execution is also vital for larger ordered write requests.
Even for 64~KB write, the throughput of \horae is half of \sysname, as more than 30\% CPU cycles are consumed for the control path.
The CPU inefficiency thus leads to the suboptimal throughput.

\subsubsection{Performance with varying batch sizes}\label{sec:eval_block_device_batch_size}
\begin{figure}[t] 
	\centering 
	\includegraphics[width=\linewidth]{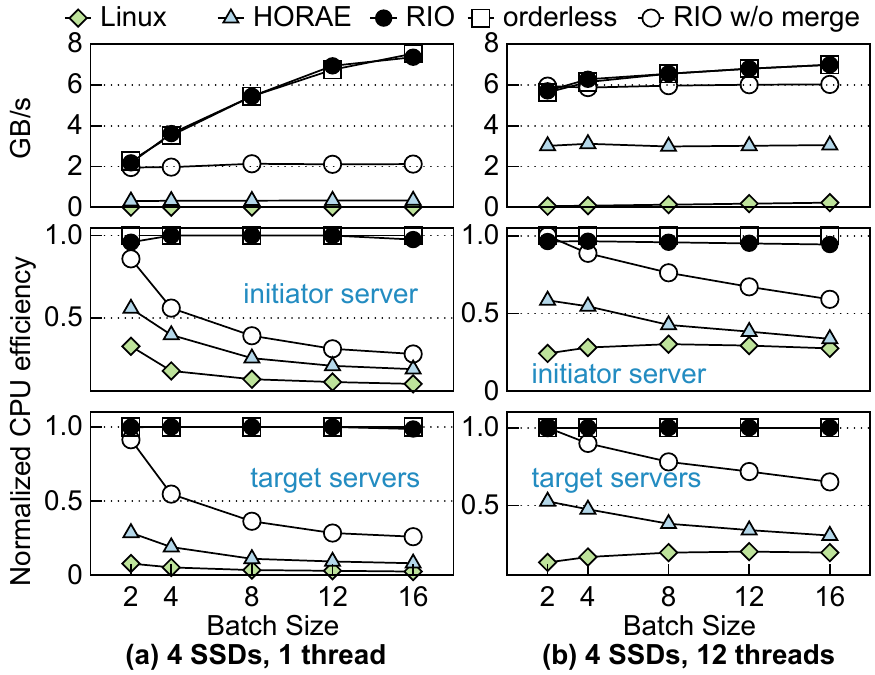} 
	\vspace{-18pt}
	\caption{\textbf{Performance with varying batch sizes.} 
	\textit{CPU efficiency is normalized to the orderless.}} 
	\label{fig:eval_bdev_batch_size} 
\end{figure}

Figure~\ref{fig:eval_bdev_batch_size} shows the throughput and CPU efficiency with varying batch sizes.
Each batch contains a sequence of 4~KB sequential write requests that can be merged.

When the computation resources are limited (Figure~\ref{fig:eval_bdev_batch_size}(a)), merging substantially reduces the CPU cycles spent on the drivers.
This further increases the overall throughput of \sysname (see the comparison between \sysname and \sysname w/o merge).
When the computation resources are sufficient (Figure~\ref{fig:eval_bdev_batch_size}(b)), as the SSDs' bandwidth is almost saturated, merging does not lead to significantly higher I/O throughput. 
Yet, \sysname retains high CPU efficiency and reserves more CPU cycles for other applications that use \sysname or \fsname.

\horae also allows merging for the data paths and merging also increases the CPU efficiency.
However, owing to the synchronous control path, increments of \horae's CPU efficiency are less significant compared to \sysname and the orderless.
Hence, the normalized CPU efficiency of \horae decreases when the batch size increases.
This indicates that the asynchronous execution at both NICs and SSDs of \sysname plays an essential role in high CPU efficiency.

\begin{figure}[t] 
	\centering 
	\includegraphics[width=\linewidth]{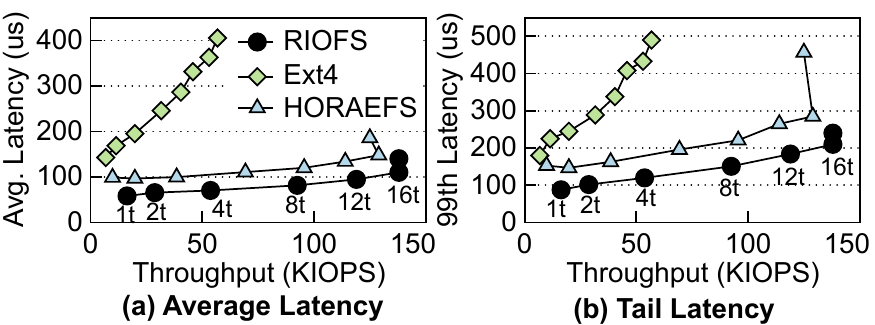} 
	\vspace{-18pt}
	\caption{\textbf{File system performance.}} 
	\label{fig:eval_fsync} 
\end{figure}

\subsection{File System Performance}\label{sec:eval_fs}
We evaluate the file system performance by \texttt{FIO}~\cite{FIO}.
The file system is mounted on the initiator and stores data on a remote Intel 905P Optane SSD.
Up to 24 threads are launched, and each issues 4~KB append \texttt{write} to a private file followed by \texttt{fsync} on the file, which always triggers metadata journaling.
Figure~\ref{fig:eval_fsync} plots the results of \texttt{fsync} calls.

\begin{figure}[t] 
	\centering 
	\includegraphics[width=\linewidth]{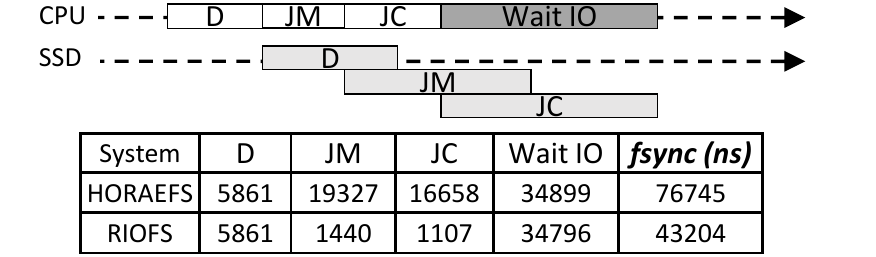} 
	\vspace{-18pt}
	\caption{\textbf{Latency breakdown.}} 
	\vspace{-10pt}
	\label{fig:eval_fsync_decompose} 
\end{figure}

We find that \sysname saturates the bandwidth of the SSD with fewer CPU cores and achieves lower latency.
Specifically, when the number of threads is 16, \sysname successfully saturates the Optane SSD.
The throughput increases by 3.0$\times$ and 1.2$\times$ in \fsname against Ext4 and \horaefs, respectively.
The average latency decreases by 67\% and 18\% in \fsname against Ext4 and \horaefs.
\fsname also makes the \texttt{fsync} less variable.
In \fsname, the 99th percentile latency decreases by 50\% and 20\% against in Ext4 and \horaefs.
The improvement of throughput and latency comes from the asynchronous execution of \sysname.
We explain this by Figure~\ref{fig:eval_fsync_decompose}.

Figure~\ref{fig:eval_fsync_decompose} presents the internal procedure of an append write (i.e., \texttt{write} followed by \texttt{fsync}) in the file system,
which consists of processing three types of data blocks: user data (D), journaled data (JM) including file system metadata and journal description block and a journal commit record (JC).
Both \fsname and \horaefs overlap the CPU and I/O processing and let the underlying device process these data blocks concurrently.
The difference lies in the way of dispatching data blocks to the lower block layer.
\horaefs leverages the control path to dispatch a group of ordered writes and experiences an extra delay.
In particular, the latency of (CPU) dispatching JM and JC increases dramatically (see the table in Figure~\ref{fig:eval_fsync_decompose}) due to the synchronous control path over the network.
\fsname can dispatch the following data blocks immediately after they reach the \textit{ORDER} queue in the block layer.
This does not need extra network round trips and thus brings performance improvement.

\subsection{Application Performance}\label{sec:eval_app}
We examine \fsname's performance with two applications, I/O intensive Varmail~\cite{filebench}, and RocksDB~\cite{rocksdb} which is both CPU and I/O intensive.
\fsname is mounted at the initiator server and stores its data on a remote Intel 905P Optane SSD.

\begin{figure}[t] 
	\centering 
	\includegraphics[width=\linewidth]{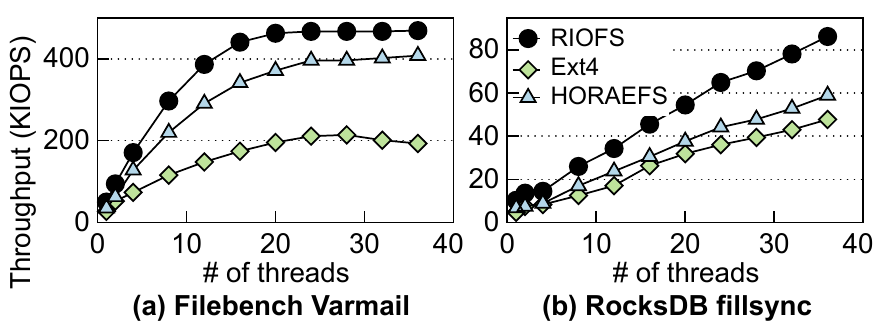} 
	\vspace{-18pt}
	\caption{\textbf{Application performance.}}
	\vspace{-18pt}
	\label{fig:app_perf} 
\end{figure}

\noindent
\textbf{Varmail.}
Varmail is a metadata and \texttt{fsync} intensive workload from Filebench~\cite{filebench}. 
We keep the default configuration and parameters of Varmail but vary the number of threads during the test.
Figure~\ref{fig:app_perf}(a) reports the results.

The throughput increases by 2.3$\times$ and 1.3$\times$ on average when we use \fsname to serve the application I/Os against when we use Ext4 and \horaefs.
Varmail contains many persistent metadata operations, e.g., \texttt{creat} and \texttt{unlink} followed by \texttt{fsync}.
\fsname provides a faster \texttt{fsync} call (details in \S\ref{sec:eval_fs}) which persists these metadata blocks in an asynchronous fashion without a serialized I/O operation as in \horaefs.
Consequently, \fsname provides higher throughput.

\noindent
\textbf{RocksDB.}
RocksDB is a popular key-value store deployed in several production clusters~\cite{rocksdb}. 
We deploy RocksDB atop the tested file systems and measure the throughput of the user requests. Here, 
we use \texttt{db\_bench}, a benchmark tool from RocksDB to evaluate the file system performance under the \texttt{fillsync} workload, which represents the random write dominant case. 
During the test, the benchmark launches up to 36 threads, and each issues 16-byte keys and 1024-byte values to a 20 GB dataset. 
Figure~\ref{fig:app_perf}(b) shows the results.

\fsname increases the throughput of RocksDB by 1.9$\times$ and 1.5$\times$ on average compared to Ext4 and \horaefs, respectively.
The performance improvement comes from two aspects: higher I/O utilization and CPU efficiency of \sysname, as we have shown in \S\ref{sec:eval_block_device}.
\fsname makes the ordered write requests asynchronous, thereby significantly increasing the I/O concurrency and reducing the CPU cycles consumed on idly waiting for block I/Os.
This in turn provides more CPU cycles for RocksDB, which also demands CPU cycles for in-memory indexing and compaction.
In the case of 36 threads, we observe that RocksDB has 110\% higher CPU utilization when we use \fsname than when we use \horaefs.
Further, \fsname packs the sequential write requests of a transaction into a larger batch (i.e., the block merging), and thus reduces CPU cycles spent on the RDMA operations over the network.
As a result, \fsname shows better performance on both CPU and I/O intensive RocksDB.

\subsection{Recovery Time}\label{sec:eval_recovery}
The recovery time of \sysname and \horae depends on the number of in-progress ordered requests before a sudden system crash.
The Linux block layer does not need to perform recovery as it permits only one in-progress ordered write request.
Here, we examine the worst case of \sysname's and \horae's recovery.
Specifically, the test launches 36 threads, and each issues 4~KB ordered write requests continuously without explicitly waiting for the completion via \texttt{fsync}.
Two target servers and 4 SSDs are used in the test.
Another thread randomly injects an error into target servers, which crashes the target driver and stops the storage service.
Then, the initiator server starts recovery after it reconnects to the target servers.
We repeat the tests 30 times and report the average results.

\sysname takes around 55~ms to reconstruct the global order, most of which is spent on reading data from PMR and transferring ordering attributes over the network.
\horae takes less time (38~ms) to reload the ordering metadata as the size of the ordering metadata is smaller than that of the ordering attribute.
The data recovery costs around 125~ms in \sysname and 101~ms in \horae, which is used for discarding the data block that disobeys the storage order.
Compared to \horae, \sysname takes more time for data recovery as the number of out-of-order requests in \sysname is higher than that in \horae.
Fortunately, discarding is performed asynchronously for each SSD and each server, and thus \sysname can fully exploit SSDs bandwidth and saves recovery time.



\section{Conclusion}\label{sec:conclusion}
This paper presents the design, implementation and evaluation of \sysname, an order-preserving networked storage stack.
By allowing ordered writes to be processed asynchronously and using a set of order-preserving techniques to enforce the persistence order, \sysname successfully drives storage devices over multiple target servers while ensuring storage order.
We conclude this work with two key observations.
First, the I/O stack should exploit the asynchronous interfaces (i.e., multiple deep hardware queues and asynchronous DMA engines) of modern NICs and SSDs to take full advantage of their high bandwidth.
Second, although block merging is expensive for the local I/O stack on ultra-low latency SSDs,
it's worth investing some CPU cycles in block merging to substantially reduce the control operations (e.g., RDMA SEND) over the network and further improve the CPU and I/O efficiency.

\section*{Acknowledgements}
We sincerely thank our shepherd Xiaosong Ma and the anonymous reviewers for their valuable feedback.
This work is funded by the National Natural Science Foundation of China (Grant No.62022051, 61832011).

\bibliographystyle{ACM-Reference-Format}
\bibliography{paper}


\end{document}